\documentclass[aip,apl,amsmath,amssymb,reprint]{revtex4-1}
\usepackage{graphicx}
\usepackage{dcolumn}
\usepackage{bm}
\usepackage[utf8]{inputenc}
\usepackage[T1]{fontenc}
\usepackage{mathptmx}
\usepackage{siunitx}
\usepackage[capitalise]{cleveref}
\usepackage{xcolor}
\usepackage{enumitem}
\usepackage{float}

\makeatletter
\def\maketitle{
	\@author@finish
	\title@column\titleblock@produce
	\suppressfloats[t]}
\makeatother

\begin{document}
	\preprint{AIP/123-QED}
	\title[]{Method of Mechanical Exfoliation of Bismuth with Micro-Trench Structures}
	\author{Oulin Yu}
	\affiliation{Department of Physics, McGill University, Montréal, Québec, H3A 2A7, Canada.}
	\author{Raphaela Allgayer}
	\affiliation{\small\textit{Department of Mining and Materials Engineering, McGill University, Montréal, Québec, H3A 2A7, Canada}}
	\author{Simon Godin}
	\affiliation{Department of Physics, McGill University, Montréal, Québec, H3A 2A7, Canada.}
	\author{Jacob Lalande}
	\affiliation{Department of Physics, McGill University, Montréal, Québec, H3A 2A7, Canada.}
	\author{Paolo Fossati}
	\affiliation{Department of Physics, McGill University, Montréal, Québec, H3A 2A7, Canada.}
	\author{Chunwei Hsu}
	\affiliation{Department of Physics, McGill University, Montréal, Québec, H3A 2A7, Canada.}
	\author{Thomas Szkopek}
	\affiliation{Department of Electrical and Computer Engineering, McGill University, Montréal, Québec, H3A 2A7, Canada.}
	\author{Guillaume Gervais}
	\email{gervais@physics.mcgill.ca}
	\affiliation{Department of Physics, McGill University, Montréal, Québec, H3A 2A7, Canada.}%
	\date{\today}
	\begin{abstract}
		The discovery of graphene led to a burst in search for 2D materials originating from layered atomic crystals coupled by van der Waals force. While bulk bismuth crystals share this layered crystal structure, unlike other group V members of the periodic table, its interlayer bonds are stronger such that traditional mechanical cleavage and exfoliation techniques have shown to be inefficient. In this work, we present a novel mechanical cleavage method for exfoliating bismuth by utilizing the stress concentration effect induced by micro-trench SiO$_2$ structures. As a result, the exfoliated bismuth flakes can achieve thicknesses down to the sub-10 nm range which are analyzed by AFM and Raman spectroscopy.
	\end{abstract}
	\maketitle
		
	\section{Introduction}
	
	\begin{figure}[!b]
		\centering
		\includegraphics[width=0.45\textwidth]{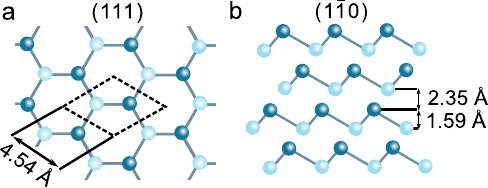}
		\caption{Layered bismuth crystal: (a) top view  of ($111$) and (b) its side view, equivalent to ($1\bar{1}0$). Note that different color shadings are used to highlight the buckled honeycomb structure~\cite{bismuthHofmann}.}\label{fig:1}
	\end{figure}
	
	\par Since the discovery of graphene earlier this century~\cite{grapheneFET}, two dimensional (2D) materials have been intensively studied and have become a topic of broad and current interest.  In 2014, black phosphorus, a group V member, was successfully exfoliated and as such is the second elemental allotrope that can be exfoliated mechanically~\cite{phosphorene1,phosphorene2,phosphorene3,phosphorene4,phosphorene5}. Similar to phosphorus, bismuth is also a member of the group V column and shares a similar layered crystal structure, as shown in \cref{fig:1}. In particular, bismuth has attracted more attention in solid state physics, as it displays noteworthy features, such as extremely large diamagnetism~\cite{curie}, high Seebeck coefficient~\cite{seebeck,ettinghausenNernst} and outstanding electronic transport properties~\cite{bismuthTransport,bismuthHofmann,bismuthAllen,bismuthGonze}. Most notably, the Shubnikov-de Haas oscillation and de Haas-van Alphen effect were also first observed in bismuth~\cite{sdh,deHaasvanAlphen}. However, most of these interesting properties have been only explored in bulk bismuth before the successful growth of single-layer bismuthene on SiC substrate in 2017~\cite{bismuthenefirst}, and the free-standing unsupported bismuthene with the buckled honeycomb structure was only grown in 2020~\cite{bismuthenefirst2}. Such a breakthrough opens doors for experimenting the electronic transport properties of 2D bismuthene since it has been predicted and shown to manifest new properties such as being a higher order topological insulator (HOTI)~\cite{hoti1,hoti2} and to host intrinsic superconductivity at temperatures below $T_c\approx0.5$ mK~\cite{sc}.
	
	\par Between its bulk form and its 2D form, bismuth thin film demonstrates tunable band gap due to the quantum confinement effects~\cite{bismuthene,2Dbeyondphosphorene}. It has been theoretically predicted and experimentally proved that, a transition from the semimetal phase with negative gap to the semiconductor phase with positive gap occurs as bismuth film's thickness reaches smaller than $\sim30\si{~\nano\meter}$~\cite{quantumEffectSsandomirskii,smsc,ultrathinBi}. In spite of these advances, as of today no transport measurements have been carefully performed in the thin film regime because of the difficulties in fabrication such as oxidization and small size of the exfoliated flakes.
	
	\par To obtain bismuth thin flakes, one would naturally expect to exfoliate bismuth in a similar fashion as its partners in the group V given that they share similar crystal structures. Unfortunately, conventional mechanical exfoliation/cleavage fails for the bismuth samples, because the van der Waals (vdW) force increases with increasing atomic weight. This is exemplified by He-He (0.0218 kcal/mol) versus Xe-Xe (0.5614 kcal/mol) vdW interactions~\cite{vdW}. Furthermore, the Bi-Bi covalent bond (197 kJ/mol) is weaker than the P-P covalent bond (490 kJ/mol)~\cite{lange}, and combined with the increasing vdW force, the anisotropy in bond strength is reduced which makes the exfoliation process difficult. Indeed, almost all the high quality samples have been prepared through molecular beam epitaxy (MBE)~\cite{walker,nagao1,nagao2,nagao3,TI1,TI2}, but MBE growth requires significantly more complex infrastructure than exfoliation based methods. Conversely, recently reported exfoliation techniques for bismuth use liquid sonication which is detrimental to the electronic properties due to high humidity and oxidation~\cite{bismutheneliquid1,bismutheneliquid2}.
	
	\par In this work, we improve the traditional mechanical cleavage technique (``drawing with chalk on a blackboard''~\cite{kim} or the scotch tape method~\cite{grapheneFET}) by fabricating micro-trench structures used as stress concentration area to break the bonds between bismuth atomic layers. The principle of this process is similar to a ``cheese grater'', where flakes are broken off as the bulk crystal is pushed against the micro-trench structures. Our experiment demonstrates that this exfoliation technique can be used to obtain ultra-thin bismuth flakes as thin as 2 nm. 
	
	\begin{figure}[!t]
		\centering
		\includegraphics[scale=0.9]{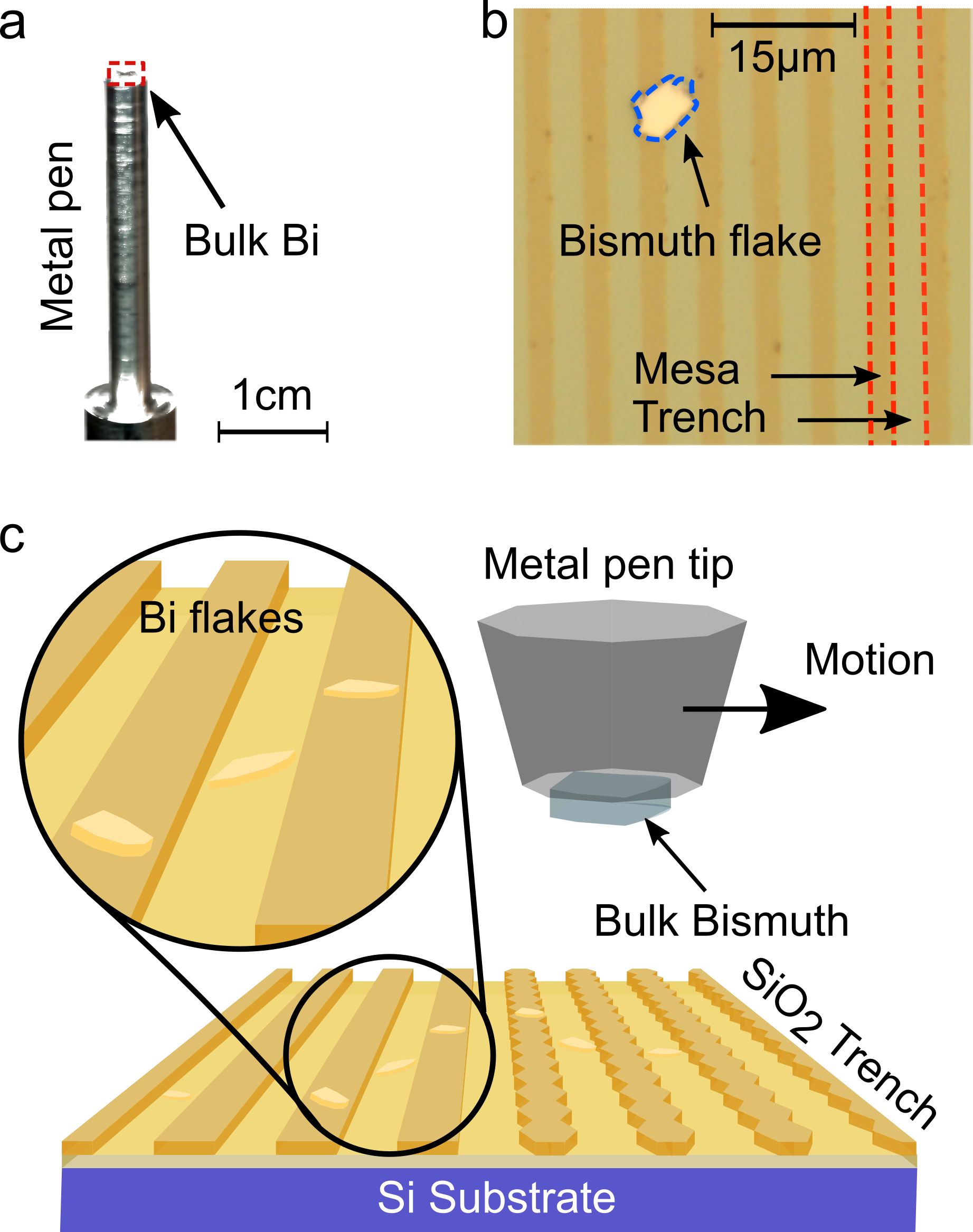}
		\caption{(a) Photograph of the metal pen with a bulk bismuth single crystal sample glued to its tip. (b) Photograph of the micro-trench structure on the SiO$_2$ substrate. The exfoliated flake is traced in blue and the red lines indicate the trench and mesa. (c) Schematic drawing which describes the exfoliation process: the metal pen holding the bismuth sample is scratched against the hard SiO$_2$ micro-trench structure in a way to break the bismuth into thin flakes.}\label{fig:2}
	\end{figure}
	
	\section{Exfoliation Method}
	
	\par As mentioned, to exfoliate bismuth, it is necessary to mechanically break the strong bonds between its atomic layers. To break the interlayer van der Waals bonds, we utilize the micro-structures to locally induce a stress concentration which then produces interlayer fractures in the bismuth. The fabricated device used to perform such exfoliation task is hence composed of two parts: a micro-trenched SiO$_2$ ``file'' or ``grater'', and a metal pen used to hold the bulk bismuth crystal. The latter is scratched against the file through pressure applied by the metal pen as {depicted} in \cref{fig:2}. Note that the hardness of bismuth, 2.25 Mohs~\cite{samsonov}, is significantly lower than that of silica, 7.0 Mohs, enabling SiO$_2$ to be an effective medium to scratch bismuth.

	\begin{figure}[]
		\centering
		\includegraphics[scale=0.8]{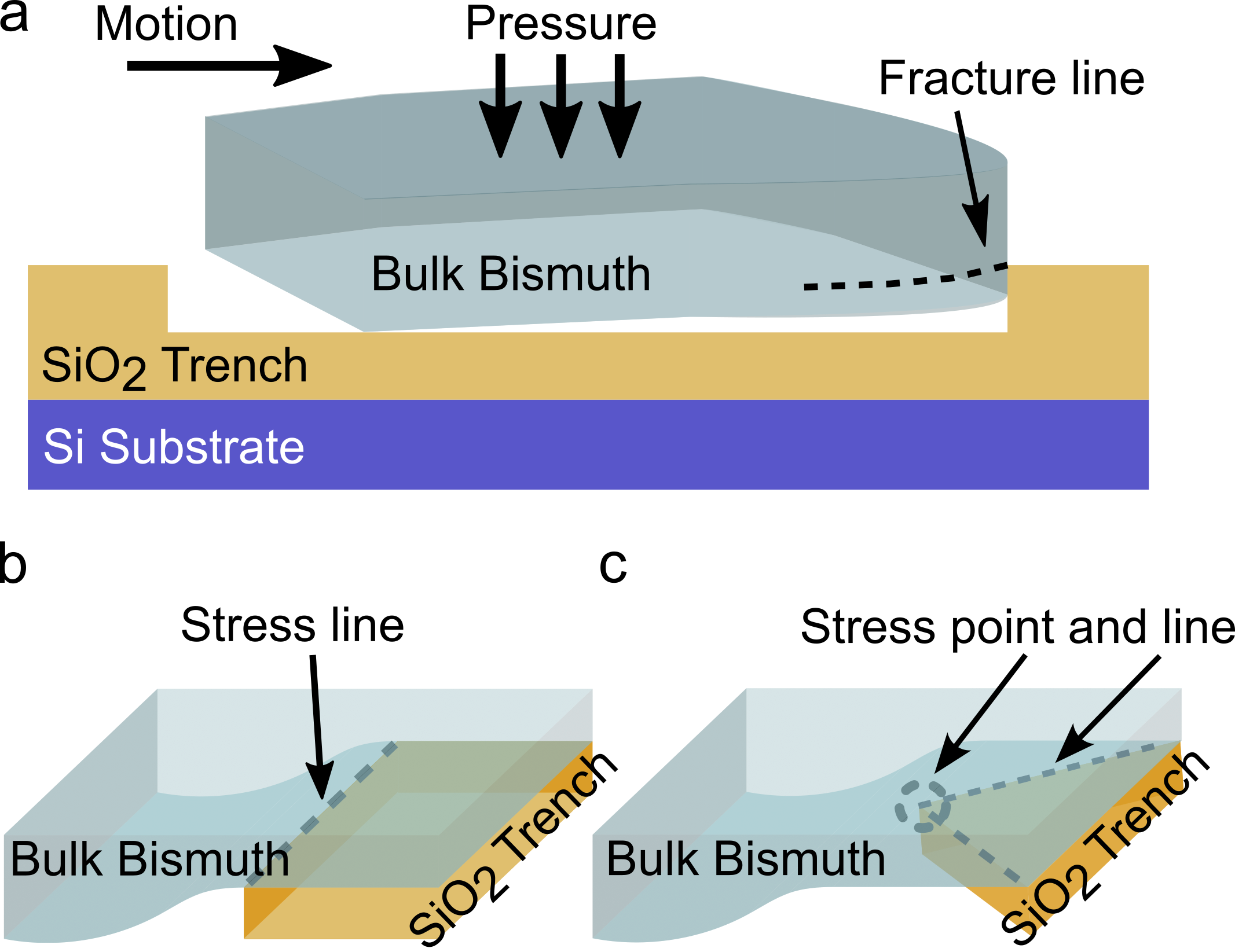}
		\caption{(a) Schematic of the bismuth crystal scratching against the SiO$_2$ micro-trench structure. Microscopically, the stress is concentrated near the trench edge and fractures the bulk crystal. (b) Bismuth crystal pressed against a straight-edge trench which results in a stress concentration line. (c) Bismuth crystal pressed against a jagged-edge trench which results in a stress concentration point.}\label{fig:3}
	\end{figure}
	
	\par A pen was fabricated to hold a small bismuth crystal at its tip. It is important to note that we prepare the bismuth crystal in a way to carefully to control the orientation of the exfoliated flakes. This is achieved by cutting the bismuth sample from a larger bismuth single crystal while ensuring the top and bottom surfaces are both $(111)$ cleavage planes. The small sample is then glued to the tip of the metal pen with a thin layer of epoxy glue. With this pen-shaped sample holder, we can conveniently control the contact pressure and the scratching speed during the exfoliation process. In our case, the bulk bismuth has a contact area of 3 mm by 3 mm and the force applied by hand is estimated to be 10 to 30 N, yielding a mean pressure of 1 to 3 MPa, which can be increased in localized regions by the micro-trenches and bismuth surface topography. {To ensure the exfoliated flakes are free of oxide, the entire process of the exfoliation is performed in a nitrogen-filled glovebox (with water and O$_2$ concentrations less than 1 ppm). Atomic force microscopy (AFM) was carried out in this nitrogen environment, and the substrate was spincoated with polymethylmethacrylate (PMMA) before exposing the flakes to air for Raman spectroscopy in order to prevent possible oxidation. The details of these characterizations are discussed later in this article.}
	
	\begin{figure*}[!ht]
		\centering
		\includegraphics[width=0.8\textwidth]{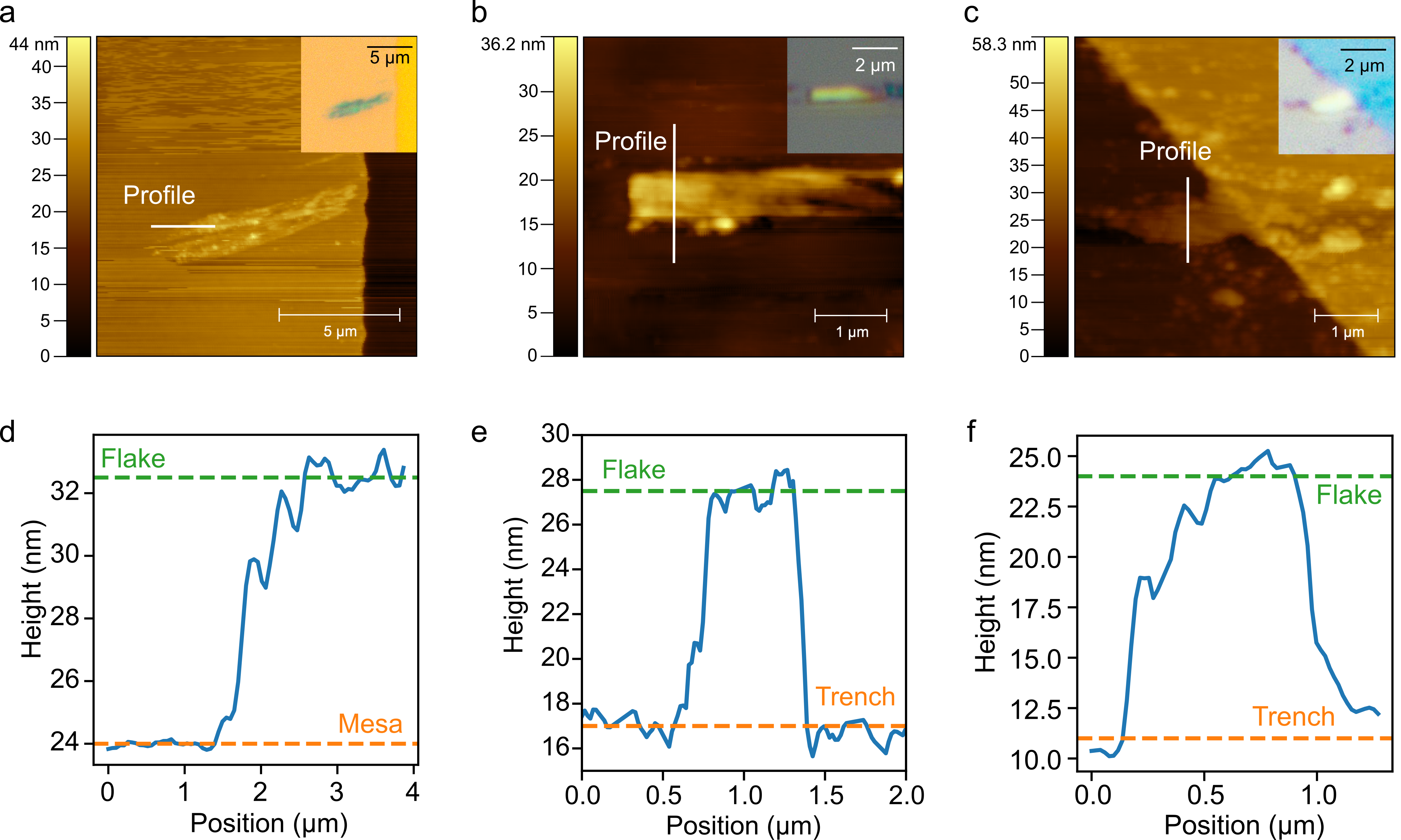}
		{\caption{(a,b,c) AFM scans of ultra-thin bismuth flakes of average thickness $4\pm2$ nm, $7\pm4$ nm and $12\pm 4$ nm. The optical images after contrast, balance and color level adjustments are shown in the insets. (d,e,f) AFM profile of exfoliated flakes along the line scans indicated in the respective AFM images.}}\label{fig:5}
	\end{figure*}
	
	\par An optical microscope image of the micro-trench structures on the SiO$_2$ file and an exfoliated flake are shown in \cref{fig:2}(b). The mesa and trench part are shown between the dotted red lines. To form these trenched structures, a 425 \si{\micro\meter}  Si wafer with 300 nm SiO$_2$ thermal oxidation layer was used. The trenches were then patterned with conventional photolithography on the oxide layer using {Microposit\textregistered} S1813 positive photoresist spun at 4000 RPM for 30 s. With the developed photoresist as a mask, the selected region of the oxide layer was partially removed with reactive ion etching (RIE). The recipe used was 100 mTorr pressure with CHF$_3$ at 45 sccm, CF$_4$ at 7 sccm, Ar at 70 sccm and power of 720 W for 5 s, yielding a trench of 25 nm deep. The remaining photoresist was removed with {Microposit\textregistered} Remover 1165, and the micro-trench structures are then completed. 
	In our experiments, we designed two types of trenches: straight-edge and jagged-edge, both shown in the schematic in \cref{fig:2}(c). The depths of these trenches are fixed at 25 nm and their widths vary from 3 $\si{\micro\meter}$ to 100 $\si{\micro\meter}$. 
	
	\par Figure 2(c) also summarizes the entire exfoliation process. The key feature of the trenched micro-structures for bismuth exfoliation is the stress concentration effect. As the SiO$_2$ is much harder than bismuth, the bismuth surface in contact with the trenched micro-structures is deformed as the metal pen is pushed against the fabricated file with appropriate pressure, and the area contacting the trench edge then becomes a stress concentrated region. \cref{fig:3}(b) and \cref{fig:3}(c) illustrate this described mechanical process for the straight-edge and jagged-edge types of trenches respectively. It is important to note that the jagged-edge trenches can induce higher stress concentration on the bismuth surface than the straight-edge trenches. This is because the sharp corners of the jagged-edge trenches can concentrate the stress to a point in addition to the stress concentration lines of the straight-edge trenches. Finally, as illustrated in \cref{fig:3}(a), while the bismuth held by the metal pen moves across the trenches, the stress concentration lines successively scan the bismuth surface, and this induces numerous fractures near the surface. From the perspective of damage mechanics, these fractures are most likely to be formed along the $(111)$ cleavage plane, which, as we have carefully arranged, is parallel to the bismuth surface. As a consequence, the stress concentrates in the fractures, and hence thin layers of bismuth flakes are stripped away from the bulk sample, leaving a trace of bismuth films with thicknesses varying from 10 nanometers to several hundred nanometers. This process is analogous to writing with a carbon lead pencil, except the lead is bulk bismuth crystal instead of graphite and the paper has a micro-trench file.
	
	\section{Result and Discussion}
	
	\par Per the exfoliation method described above, we succeeded to produce bismuth flakes of various sizes and thicknesses. While our method is effective for producing thin bismuth flakes, we observed no difference in the performance of the straight-edge and jagged-edge types of trenches. We attribute this observation to the fact that the edges only introduce fractures to the bulk crystal and do not shave the crystal directly. Furthermore, the exfoliated flakes do not prefer to fall into the trenches and are observed in the trenches as well as on the mesas. This  also affirms that fractures are only introduced by the micro-trench structures and that the actual exfoliation of flakes is responsible by the ``writing'' process. \cref{fig:2}(b) shows the optical microscope image of a typical flake on the  SiO$_2$ file. Note that this flake has a height over 100 nm giving a good visibility, and thus can be easily identified with an optical microscope. In the \textit{Supplemental Material}, we calculate the optimal filter required to maximize the contrast of a sub-10 nm bismuth flake on the micro-trench structures.
	
	\par Ultra-thin bismuth flakes were also obtained with our exfoliation technique. {Figure 4 shows three different flakes of average thickness $4\pm2$ nm, $7\pm4$ nm and $12\pm 4$ nm. The average thickness is obtained by averaging the height of the flake and subtracting the average substrate height. The AFM scans are shown in \cref{fig:5}(a), (b) and (c), and a profile is selected and plotted in \cref{fig:5}(d), (e) and (f).} Their optical microscopic images taken at 100$\times$ magnification are shown in insets of \cref{fig:5}(a), (b) and (c). Note that given the poor optical contrast of these ultra-thin flakes, the colors in these images are adjusted to enhance the visibility of these flakes by tuning the contrast, white balance and color levels. 
	\begin{figure}[!ht]	
		\centering
		\includegraphics[scale=0.7]{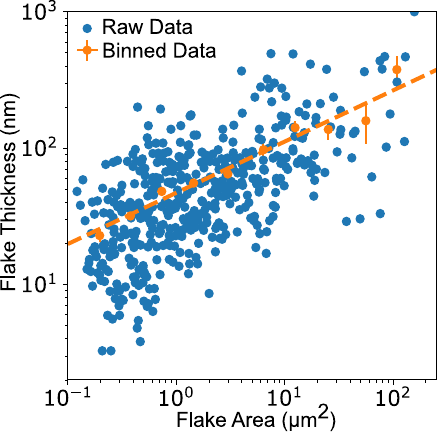}
		\caption{Flake height versus flake area in log-log scale. Binned (logarithmically) data with error bars given by the standard error. The fit reveals that $\text{thickness}\propto\text{area}^{0.38}$.}\label{fig:6}
	\end{figure}
	
	\par As to other types of exfoliation techniques, our method can produce flakes with various sizes and thicknesses. We have analyzed AFM images of 469 flakes and recorded their height and surface area. To better reveal the trend, these data were binned logarithmically by area and are shown on a logarithmic scale in {\cref{fig:6}}. A clear linear trend is observed on this log-log scale, suggesting a power-law relationship $\propto \text{ area}^{0.38}$. As the data demonstrates, the flake thickness decreases as the surface area decreases, however, it is possible to obtain large area flakes with relatively small thicknesses in the sub-100 nm range. To fabricate a four terminal device for exploring electrical transport properties, a flake area of 1 \si{\micro\meter\squared} would be necessary. As it is seen in \cref{fig:5}, for an area larger than 1 \si{\micro\meter\squared}, we can obtain flakes as thin as 10 nm and often in the tens of nanometer range which is ideal to explore, among other predictions, the semi-metal to semi-conductor transition of bismuth~\cite{smsc}.
		
	\begin{figure}[ht!]
		\centering
		\includegraphics[scale=0.6]{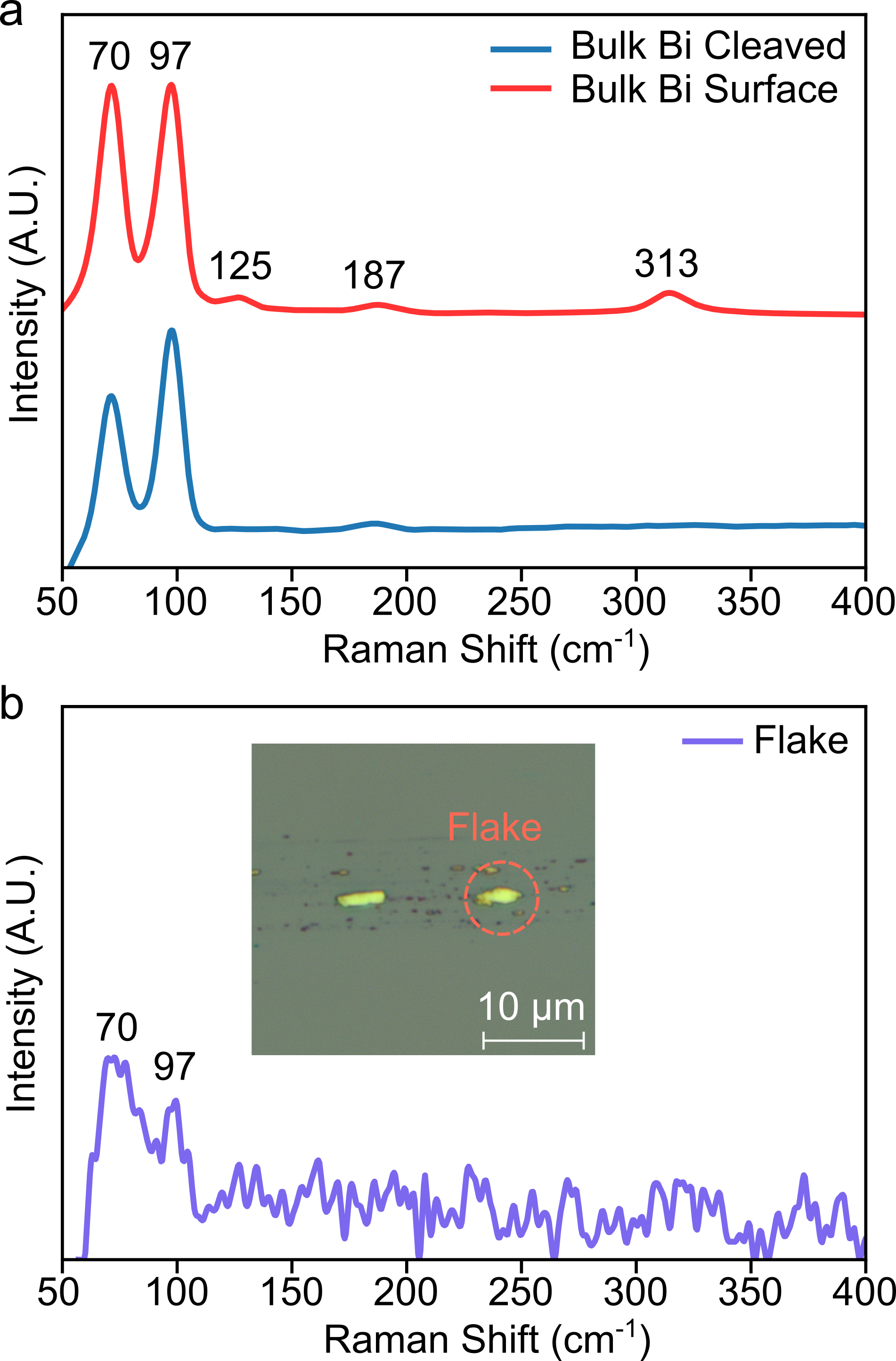}
		\caption{Raman spectroscopy of the bulk bismuth. The bismuth Raman peaks at 70 cm$^{-1}$ and 97 cm$^{-1}$, and the $\beta$-phase of bismuth oxide's Raman peak at 313 cm$^{-1}$ are shown.}\label{fig:bulkraman}
	\end{figure}
	
	\section{Raman Characterization}

	\par Raman spectroscopy was performed in air on exfoliated flakes protected by a  polymethylmethacrylate (PMMA) cap layer that was spun coated in a nitrogen glove box environment, as well as on the bulk crystal for benchmarking.  The laser wavelength is 785 nm with 50$\times$ objective and 25 \si{\micro\meter} aperture diameter. \cref{fig:bulkraman}(a) shows the Raman spectra of the bulk crystal's surface before and after it was freshly cleaved. Both bismuth Raman shift peaks at 70 cm$^{-1}$ and 97 cm$^{-1}$ for the $E_g$ and $A_{1g}$ vibrational modes, and as well as the overtone at 187 cm$^{-1}$, are clearly observed~\cite{walker,ramanBi}, as expected for bismuth crystal. However, only the Raman spectrum on the surface of the bulk crystal clearly demonstrates the characteristic Raman shifts at 125 cm$^{-1}$ and 313 cm$^{-1}$ for the Bi-O bonds in $\beta$-phase bismuth oxide where the latter is commonly formed at lower temperatures under the 300$^\circ$C range~\cite{ramanBiOxide1,ramanBiOxide2,ramanBiOxide3}. In contrast, after cleaving the bulk bismuth specimen and exposing its interior, the bismuth oxide peak disappeared, indicating that the oxidation is only present on the surface of the bulk crystal. Consequently, our mechanical exfoliation process is done in a glovebox, provoding a nitrogen-only environment with O$_2$ and water concentrations less than 1 ppm, hence avoiding oxidation of the exfoliated thin bismuth flakes.
	
	\par To confirm this, a small flake of thickness 160 nm obtained from grating the bulk crystal against micro-trench structures was also analyzed with Raman spectroscopy where the Raman shifts is shown in \cref{fig:bulkraman}(b) along with the optical image of the flake in the inset. Given the small thickness and area of the flake, its signal is rather small. Nevertheless, the Raman peaks for bismuth at 70 cm$^{-1}$ and 97 cm$^{-1}$ are still observed unambiguously, confirming the presence of bismuth in the exfoliated flakes. The Raman peaks for $\beta$-phase bismuth oxide could not be observed within the detection limit of our Raman apparatus. This indicates that there is little or no sign of oxidation of the exfoliated flake. Indeed, given the small size of the flake, we expect the oxidation to be throughout the flake instead of only on its surface, hence the oxide peaks would dominate the Raman spectrum.

	\section{Conclusion}
	\label{sec:conclusion}
	\par In this work, we described a novel exfoliation technique based on stress concentration effects induced by micro-trench structures. Previous mechanical cleavage methods use adhesion to a flat surface to exfoliate a layered crystal, or take advantage of the weak van der Waals force between atomic layers where bulk crystals are rubbed against a solid surface as in ``drawing with chalk''. In contrast, our approach uses the shear stress produced by scratching the bulk crystal against the hard SiO$_2$ substrate with micro-trench structures, acting as a ``cheese grater''. Such shearing motion produces stress concentration in the bulk crystal in contact with the trench edge, and consequently fractures are produced along the stress concentration line/point to strip thin layers. 
	
	\par We have shown that flakes as thin as $\sim$10 nm can be obtained with this exfoliation method, albeit they become barely visible under an optical microscope. In the \textit{Supplemental Material}, we studied the contrast for ultra-thin bismuth flakes and showed that a red filter between the range of 500 - 570 nm greatly improves the flake visibility in the sub-10 nm range. 
	
	\par Most importantly, our work demonstrates the concept of using stress concentration to mechanically exfoliate crystals with strong bonds between interlayers. In the future, this technique could prove useful to fabricate devices with 2D materials thought not possible to exfoliate via traditional mechanical exfoliation method.
	
	\section*{Supplemental Material}
	\par See supplemental material for the theoretical contrast optimization.
	
	\begin{acknowledgments}
		\par This work has been supported by the New Frontiers in Research Fund program, the Natural Sciences and Engineering Research Council of Canada, the Canadian Institute for Advanced Research, the Canadian Foundation for Innovation, and the Fonds de Recherche du Québec Nature et Technologies. Sample fabrication was carried out at the McGill Nanotools Microfabrication facility and GCM Lab at Polytechnique de Montréal. We would like to thank R. Talbot, R. Gagnon, and J. Smeros for technical assistance and M.-H. Bernier for helpful discussions. Finally, we would like to thank Qian Luo for his many contributions to this work.
	\end{acknowledgments}

\clearpage
\newpage\pagebreak

\title[]{Supplemental Material: Method of Mechanical Exfoliation of Bismuth with Micro-Trench Structures}
\maketitle
\onecolumngrid
\section*{Contrast Optimization}

\begin{figure}[H]
	\includegraphics[scale=1]{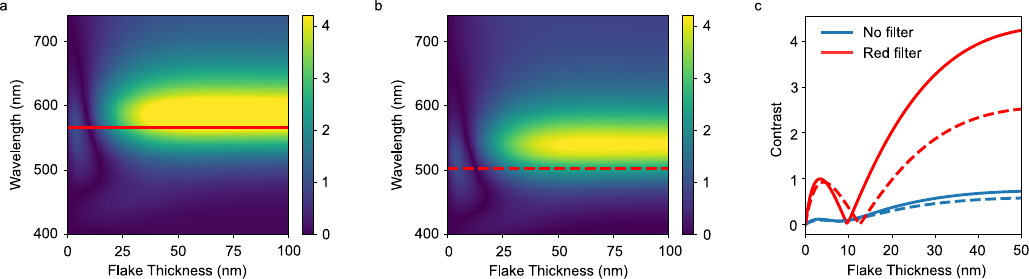}
	\caption{Contrast plots for (a) 300 nm silicon oxide (mesa) and (b) 275 nm silicon oxide (trench). The red lines are the optimized wavelengths (567 nm for mesa and 503 nm for trench) for ultra-thin flake visiblility. (c) These optimal wavelengths are extracted and compared to the contrast given by the optical microscope's halogen bulb (blue line is for mesa, and dashed blue line is for trench). }\label{fig:4}
\end{figure}

\par As noted in the main text, thin flakes become more difficult to observe as their thickness decreases. Since flakes are identified visually through an optical microscope, one can easily overlook flakes that are ultra-thin. To mitigate this, we studied the contrast for ultra-thin bismuth flakes on SiO$_2$ micro-trench structures following the work done by Blake \textit{et al.}  for graphene~\cite{blake}.
\par The contrast is defined as the normalized absolute difference between reflected light intensity in the presence and absence of a bismuth flake, specifically 
$$C = \left|\dfrac{R-R_0}{R_0}\right|,$$
where $C$ is the contrast, $R_0$ is the reflectivity in the absence of a bismuth flake and $R$ is the reflectivity in the presence of a bismuth flake. We show in \cref{fig:4}(a) and \cref{fig:4}(b)  the contrast colorplots for the mesa (300 nm SiO$_2$) and the trench (275 nm SiO$_2$), respectively. We chose to use the indices of refraction as in Palik~\cite{palik} for Si and SiO$_2$ as well as Hagemann \textit{et al.}~\cite{hagemann} for bismuth. However we also compared to indices of refraction reported by Lenham \textit{et al.}~\cite{lenham} and by Abu El-Haija \textit{et al.}~\cite{abu}, and found negligible difference for the ultra-thin bismuth flakes analyzed here.
\par In both \cref{fig:4}(a) and (b), the bright yellow part shows high contrast for flakes larger than 25 nm in the optical wavelength range, however for flake thicknesses less than 20nm, the contrast decreases drastically. This makes the ultra-thin bismuth flakes very difficult to observe with the naked eye and often requires AFM imaging to confirm the presence of bismuth flake.
\par This can be improved by using filters. Based on the our calculations, the contrast for sub-10 nm bismuth flakes can be optimized with a filter of 503 nm and 567 nm for mesa and trench respectively, which corresponds to a red filter in the electromagnetic spectrum. This is shown in \cref{fig:4}(c) by the red lines where the contrast for these two wavelengths are compared to the contrast averaged over power spectral density of the optical microscope's halogen bulb for both the mesa (300 nm) and trench (275 nm). Consequently, the use of a red filter improves the visibility of the sub-20 nm bismuth flakes.

\end{document}